\documentclass[twocolumn,showpacs,preprintnumbers,amsmath,amssymb]{revtex4}
\usepackage{graphicx}
\usepackage{dcolumn}
\usepackage{bm}
\usepackage{epsfig}
\newcommand{\QKD}{{\sc QKD}}
\newcommand{\PNS}{{\sc PNS}}
\newcommand{\WCP}{{\sc WCP}}

\newcommand{\OA}{{\sc OA}}

\newcommand{\ket}[1]{\mbox{$ | #1 \rangle $}}
\newcommand{\bra}[1]{\mbox{$ \langle #1 | $}}
\newcommand{\si}{\sigma}
\newcommand{\one}{\leavevmode\hbox{\small1\normalsize\kern-.33em1}}

\begin{document}

\title{Practical quantum key distribution: On the security evaluation with inefficient single-photon detectors}
\author{Marcos Curty and Norbert L\"{u}tkenhaus}
\affiliation{Institut f\"ur Theoretische Physik I, 
Institut f\"ur Optik, Information und Photonik
(Max-Planck-Forschungsgruppe), Staudtstr. 7/B3, Universit\"{a}t 
Erlangen-N\"{u}rnberg, 91058 Erlangen, Germany}

\begin{abstract}
Quantum Key Distribution with the BB84 protocol has been shown to
be unconditionally secure even using weak coherent pulses instead
of single-photon signals. The distances that can be covered by
these methods are limited due to the loss in the quantum channel
(e.g. loss in the optical fiber) and in the single-photon counters
of the receivers. One can argue that the loss in the detectors
cannot be changed by an eavesdropper in order to increase the
covered distance. Here we show that the security analysis of this
scenario is not as easy as is commonly assumed, since already
two-photon processes allow eavesdropping strategies that
outperform the known photon-number splitting attack. For this
reason there is, so far, no satisfactory security analysis
available in the framework of individual attacks.
\end{abstract}
\maketitle

\section{INTRODUCTION}
Quantum key distribution (\QKD) \cite{Wiesner83,BB84} is a
technique that allows two parties (Alice and Bob) to generate a
secret key despite the computational and technological power of an
eavesdropper (Eve) who interferes with the signals. Together with
the Vernam cipher \cite{Vernam26}, \QKD\ can be used for
unconditionally secure data transmission.

The basic ingredient of any \QKD\ protocol is the distribution of
{\it effective} quantum states that can be proved to be entangled
\cite{Curty03}. The first complete scheme for \QKD\ is that
introduced by Bennett and Brassard in $1984$ (BB$84$ for short)
\cite{BB84}. In a quantum optical implementation of this protocol,
Alice encodes each random bit into the polarization state of a
single-photon. She chooses for her encoding one of two mutually
unbiased bases, e.g. either a linear or a circular polarization
basis. On the receiving side, Bob measures each photon by
selecting at random between two polarization analyzers, one for
each possible basis. Once this phase is completed, Alice and Bob
use an authenticated public channel to process their correlated
data in order to obtain a secret key. This last procedure, called
{\it key distillation}, involves, typically, postselection of
data, error correction to reconcile the data, and privacy
amplification to decouple the data from Eve \cite{Norbert99}. A
full proof of the security for the whole protocol has been given
in \cite{Mayers98,Shor00,Biham00,Lo99}.

After the first demonstration of the feasibility of this scheme
\cite{Bennett92}, several long-distance implementations have been
realized in the last years (see
\cite{Marand95,Bethune02,Stucki02,Hughes02} and references
therein). However, these practical approaches differ in many
important aspects from the original theoretical proposal, since
that demands technologies that are beyond our present experimental
capability. Especially, the signals emitted by the source, instead
of being single-photons, are usually weak coherent pulses (\WCP)
with typical average photon numbers of $0.1$ or higher. These
pulses are described by coherent states in the chosen polarization
mode. The quantum channel introduces considerable attenuation and
errors that affect the signals even when Eve is not present.
Finally, the detectors employed by the receiver have a low
detection efficiency and are noisy. All these modifications
jeopardize the security of the protocol, and leads to limitations
of rate and distance that can be covered by these techniques
\cite{Huttner95,Norbert00}. A positive security proof against all
individual particle attacks, even with practical signals, has been
given in \cite{Norbert00b}. More recently, a complete proof of the
unconditional security of this scheme in a realistic setting has
been achieved \cite{Inamori01}. This means that, despite practical
restrictions, with the support of the classical information
techniques used in the key distillation phase, it is still
possible to obtain a secure secret key.

The main limitation of \QKD\ based on \WCP\ arises from the fact
that some pulses contain more than one photon prepared in the same
polarization state. Now Eve is no longer limited by the no-cloning
theorem \cite{Wootters82} since in these events the signal itself
provides her with perfect copies of the signal photon. She can
perform the so called {\it Photon Number Splitting} (\PNS) attack
on the multi-photon pulses \cite{Huttner95}. This attack provides
Eve with full information about the part of the key generated with
the multi-photon signals \cite{NoteGisin}, without causing any
disturbance in the signal polarization. Together with an optimal
eavesdropping attack on the single-photon pulses, the \PNS\ attack
constitutes Eve's optimal strategy \cite{Norbert00b,Inamori01}.
This result is stated for a conservative definition of security.
In this paradigm, it is commonly assumed that some flaws in Alice
and Bob's devices (e.g.  the detection efficiency and the dark
count probability of the detectors), together with the losses in
the channel, are controlled by Eve, who exploits them to obtain
maximal information about the shared key.

In this paper we analyze a different scenario. We impose
constraints on Eve's capabilities, and we are interested in the
influence that this effect has on her best strategy. It is
necessary to distinguish this work from earlier ones: here we
consider a more relaxed definition of security than the one in
\cite{Norbert00b,Inamori01}. In particular, we study the situation
where Eve is not able to manipulate Alice and Bob's devices at
all, but she is limited to act exclusively on the quantum channel
(See e.g. \cite{dusek00a}). The main motivation to consider this
scenario is that from a practical point of view it constitutes a
reasonable description of  a realistic situation, where Alice and
Bob can limit Eve's influence on their apparatus by some
counterattack techniques. However, this scenario has not been
analyzed thoroughly. See  Appendix \ref{why} for a discussion of
the articles by Gilbert and Hamrick. In discussions within the
scientific community one often hears  the hope that it is
sufficient to consider the \PNS\ attack, but this time taking into
account the finite detection efficiency of Bob's detectors. As a
result the loss of a photon in the \PNS\ attack reduces the
probability to detect the remaining  signal with the inefficient
detectors and less multi-photon signals contribute to the final
key. This suggests higher available rates. However, the analysis
of this scenario is rather subtle, as we will show in this paper.
Note that a first counter example against that believe is
contained in \cite{gilbert00a} showing that the unambiguous state
discrimination attack of \cite{dusek00a} can outperform the
adaptation of the photon-number splitting attack of
\cite{Norbert00,Norbert00b} in the discussed scenario of limited
eavesdropping capabilities. This result applied to signals
containing at least three photons. We show that already two-photon
processes allow for improved eavesdropping in the restricted
scenario.

With our article, we   point out the difficulty of
analysing the scenario where Bob's detection efficiency cannot be
manipulated. For this we refer to the standard BB$84$ protocol, where
in the first part only
the raw bit rate (before the key distillation phase) is
monitored but not the number of coincidence detections. We construct two specific eavesdropping
strategies which do not subtract photons from all the
multi-photon pulses, and that are more powerful than the \PNS\
attack for some relevant regimes of the observed error rate. They
are based on specifically chosen cloning attacks. The results
obtained here do not constitute a complete analysis of Eve's
optimal attack under these restriction, they introduce a new class of
eavesdropping strategies that become relevant only in this scenario. Our results
clearly show that a simple extension of the \PNS\ attack in this
scenario fails to deliver security.  In an extended version of
the protocol, where Alice and Bob can access the complete photon
number statistics of the arriving signal, we find that the
advantage of the cloning attacks is not as evident, but requires a
deeper analysis.

The paper is organized as follows.  In Sec.~\ref{sec:TOOLS} we
describe in more detail the scenario we consider here. This
includes the signal states and detection methods employed by Alice
and Bob together with the technologies assumed for Eve. In
Sec.~\ref{sec:PNS} we introduce the complete \PNS\ attack, and we
analyze a particular process that is part of this attack and
involves only single-photon signals and two-photon signals. In
Sec.~\ref{sec:CLON} we introduce two more processes that do not
subtract photons from the pulses. They are based on cloning
machines operating only on two-photon pulses. In
Sec.~\ref{sec:COMP}, these two processes are compared with the
\PNS\ process. We show that, for some relevant regimes of the
observed error rate in the sifted key, the two processes based on
cloning machines provide Eve with more information than the \PNS\
process. The extended version of the protocol, where Alice and Bob
use the full statistics at their disposal to detect Eve, as
introduced in \cite{Norbert99,dusek00a,Norbert99b} is briefly consider in
Sec.~\ref{sec:STAT}. Finally, Sec.~\ref{sec:CONC} concludes the
paper with a summary.

\section{TOOLBOX FOR ALICE, BOB AND EVE}\label{sec:TOOLS}
\subsection{Alice}
Alice uses \WCP\ signal states that are described by coherent
states with a small amplitude $\alpha$. This corresponds to the
description of a dimmed laser pulse. We consider otherwise a
perfect implementation of the signal states. The coherent state is
given by
\begin{equation}
\ket{\alpha}=e^{-|\alpha|^2/2}\sum_{n=0}^{\infty}\frac{(\alpha{}a^\dag)^n}{n!}\ket{0},
\end{equation}
with $a^\dag$ being the creation operator for one of the four
BB$84$ polarizations modes. However, usually there is no
reference phase available outside Alice's lab, and the state that
Bob and Eve see is not a coherent state $\ket{\alpha}$, but the
phase-averaged form of the signal,
$\rho=\frac{1}{2\pi}\int_{\phi}\ket{e^{i\phi}\alpha}\bra{e^{i\phi}\alpha}\
{\mathrm d}\phi$. This results in an effective signal state which
is a mixture of Fock states with a Poissonian photon-number
distribution of mean $\mu=|\alpha|^2$. It is described by the
density matrix
\begin{equation}\label{SigAlice}
\rho=e^{-\mu}\ \sum_{n=0}^{\infty}\
\frac{\mu^n}{n!}\ket{n}\bra{n},
\end{equation}
where the state $\ket{n}$ denotes the Fock state with $n$ photons
in one of the four BB$84$ polarization states.
\subsection{Bob}
\begin{figure}
\begin{center}
\includegraphics[scale=.9]{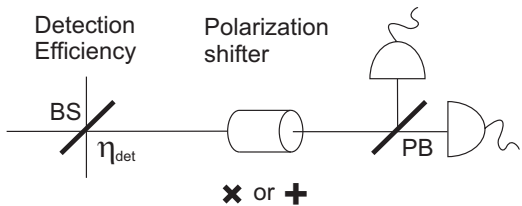}
\end{center}
\caption{The polarization shifter allows to change the
polarization basis ($+$ and $\times$) of the measurement as
desired. The polarization analyzer consists of a polarizing beam
splitter (PB) and two ideal detectors. The PB discriminates the
two orthogonal polarized modes. Detection efficiencies are modeled
by a beam splitter (BS) of transmittance $\eta_{det}$.
\label{Detector}}
\end{figure}

We consider that Bob employs the active detection setup shown in
Fig.~\ref{Detector}.
It consists of a polarization analyzer and a polarization shifter
which effectively changes the polarization basis of the subsequent
measurement. The polarization analyzer has two detectors,
monitoring each one output of a polarizing beam splitter. These
detectors are characterized by their detection efficiency
$\eta_{det}$. They can be described by a combination of beam
splitters of transmittance $\eta_{det}$ and ideal detectors
\cite{Yurke85}. This model can be simplified further by
considering that both detectors are equal. In this situation, it
is possible to attribute the losses of both detectors to a
single-loss beam splitter which is located after the transmission
channel. We assume that the detectors cannot distinguish the
number of photons of arrival signals, but they provide only two
possible outcomes: ``click" (at least one photon is detected), and
"no click" (no photon is detected in the pulse).

The action of Bob's detection device can be characterized by two
POVM, one for each of the two polarization basis $\beta$ used in
the BB$84$ protocol \cite{Note1}. Each POVM contains four elements
\cite{Norbert99b}: $F_{vac}^\beta, F_0^\beta, F_1^\beta$, and
$F_D^\beta$. The outcome of the first operator, $F_{vac}^\beta$,
corresponds to no click in the detectors, the following two POVM
operators, $F_0^\beta$ and $F_1^\beta$, give precisely one
detection click (these are the desired measurements), and the
last one, $F_D^\beta$, gives rise to both detectors being
triggered. If we denote by $\ket{n,m}_\beta$ the state which has
$n$ photons in one mode and $m$ photons in the orthogonal
polarization mode with respect to the polarization basis $\beta$,
the elements of the POVM for this basis are given by
\begin{eqnarray}\label{det}
F_{vac}^\beta&=&\sum_{n,m=0}^{\infty}\ \bar{\eta}^{n+m}\
\ket{n,m}_\beta\bra{n,m},
\\F_{0}^\beta&=&\sum_{n,m=0}^{\infty}\ (1-\bar{\eta}^n)\bar{\eta}^{m}\
\ket{n,m}_\beta\bra{n,m}, \nonumber\\
F_{1}^\beta&=&\sum_{n,m=0}^{\infty}\
(1-\bar{\eta}^m)\bar{\eta}^{n}\
\ket{n,m}_\beta\bra{n,m},\nonumber\\
F_D^\beta&=&\sum_{n,m=0}^{\infty}\
(1-\bar{\eta}^n)(1-\bar{\eta}^m)\ \ket{n,m}_\beta\bra{n,m},
\nonumber
\end{eqnarray}
where $\bar{\eta}=(1-\eta_{det})$.

The detectors show also noise in the form of dark counts which
are, to a good approximation, independent of the signal. Note
that the observed errors can be though as coming from a two-step
process: In the first step the signals are changed as they pass
Eve's domain in the quantum channel, in the second step random
noise from the detector dark counts is added. If we assume that
the second step cannot be influenced by Eve, then Alice and Bob
can infer the channel error rate, that is assumed to be due to
eavesdropping, from their data and their knowledge of the
detector performance. This means that only this reduced channel
error rate needs to be taken into account in the privacy
amplification step.
\subsection{Eve}
As discussed before, we allow Eve to have at her disposal all
technology allowed by Quantum Mechanics, but she is limited to
use it exclusively on the quantum channel. This assumption has two
consequences on Eve's possible eavesdropping strategies that are
vital for the security analysis of the next sections. In
particular, the detection efficiency $\eta_{det}$ of Bob's
detectors is fixed and Eve cannot influence it to obtain extra
information \cite{Note2}. Moreover, since we consider that the
noise of the detectors is independent of the signals entering
them, Eve cannot make use of the dark counts to increase her
information.
\section{THE PNS ATTACK}\label{sec:PNS}
In the photon-number splitting attack Eve performs a {quantum
non-demolition measurement} of the {\em total} number of photons
of each signal. Whenever she finds that a signal contains two or
more photons, she deterministically takes one photon out. The
remaining photons are then forwarded to Bob. The photons in Eve's
hand will reveal its signal polarization to Eve if she waits with
her measurement until she learns the polarization basis during the
key distillation phase.
If the loss of the channel is strong enough, Eve can block all the
single-photon pulses and forward only  the remaining photons of
multi-photon signals by a lossless channel; on these signals she
can obtain the whole information.
In this situation no secure key can
be generated. When the loss is not high enough for this, then Eve
can block only a fraction of the single-photon signals, but she
can perform some optimal eavesdropping attack on the remaining
single-photon pulses. Moreover, the whole process can be adapted
such that it mimics the photon number statistics of a lossy
channel in typical situations \cite{Norbert02}.

When Bob uses a detection setup with ideal detectors, or Eve can
manipulate their efficiency such as $\eta_{det}=1$, then the \PNS\
attack constitutes Eve's optimal strategy
\cite{Norbert00b,Inamori01}. The reason is that in this case all
 signals that provide Eve with full information about the key
(multi-photon pulses) contribute for the raw key. If the detectors
have a detection efficiency $\eta_{det}<1$ which Eve cannot
change, we find that with certain probability the multi-photon
signals can also contribute to vacuum events in the detection
process. In this situation, there are  regimes where the \PNS\
attack is still Eve's optimal eavesdropping strategy. This happens
when the loss in the channel is sufficiently high such that the
number of non-vacuum signals expected to {\it arrive} at Bob's
detection device is smaller than the number of multi-photon
signals. Here we consider the regime where this is not the case.
This means that Eve needs to compensate the effect of the
undetected multi-photon signals by increasing the number of
single-photons signals that is sent to Bob. This fact reduces the
effectiveness of the \PNS\ attack, and one might consider the
existence of better strategies for Eve.

\begin{figure}
\begin{center}
\includegraphics[scale=.7]{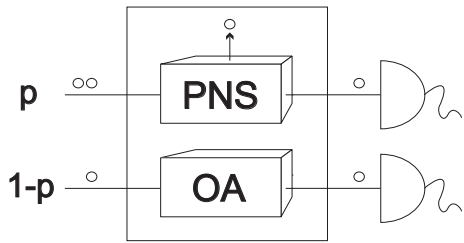}
\end{center}
\caption{Process included in the \PNS\ attack. With probability
$p$ the pulse contains two photons, Eve takes on photon out of
it, and she sends the remained one photon to Bob. In the case of
single-photon signals (probability $1-p$), Eve performs an optimal
eavesdropping attack (\OA) on these pulses. \label{PNS}}
\end{figure}

We focus in a particular combination of processes that are
contained in the extended \PNS\ attack \cite{Norbert02}, now with
imperfect detectors. This combination includes only some
two-photon processes (with probability $p$) and some one-photon
processes (probability $1-p$) from the whole eavesdropping
strategy. It is represented in Fig.~\ref{PNS}.
The objective is to obtain Eve's maximum information on this
combination of processes given a particular disturbance in the
signals. For that, we employ the concept of mutual information
given by Shannon. Under this definition, it has been proven that
the optimal attack on single-photon signals (\OA), i. e. the one
that provides Eve maximum information about the raw key, coincides
with the optimal individual attack on these signals
\cite{Xiangbin01}. This optimal individual attack has been
introduced by Fuchs {\it et. al.} in \cite{Fuchs97}. In the
symmetric strategy, every qubit signal $\rho_A$ sent by Alice is
transformed into the mixed state $\rho_B=(1-2D)\rho_A+D\one$. The
disturbance $D$ represents the error rate in the sifted key within
the chosen signals; it is not the overall observed error rate. The
connection between the two error rates is made in section
\ref{observed}. For a given value of $D$, Eve's maximum
information in this attack is given by \cite{Fuchs97}
\begin{equation}
\label{IAEopt} I_{AE}=\frac{1}{2}\Phi\Big(2\sqrt{D(1-D)}\Big),
\end{equation}
where the function $\Phi$ is defined as
$\Phi(x)=(1+x)\log_2(1+x)+(1-x)\log_2(1-x)$. With this result,
now it is straightforward to obtain Eve's maximum information in
the \PNS\ process of Fig.~\ref{PNS}, as a function of $p$ and $D$,
\begin{equation}\label{eqPNS}
I_{AE}^{PNS}=p+\frac{1-p}{2}\Phi\Big(2\sqrt{D(1-D)}\Big).
\end{equation}

In the next section we introduce two more combination of processes
that have the same input signals as those of Fig.~\ref{PNS}. Then,
in Sec.~\ref{sec:COMP} we show that these new processes provide
Eve more information than the \PNS\ process, for some relevant
regimes of $D$. Moreover, the raw bit rate of all the processes
can be selected to be the same. This means that the substitution
of the combination of processes of Fig.~\ref{PNS} by any of the
new combinations leads to a better eavesdropping strategy in terms
of Shannon information.

\section{CLONING ATTACKS}\label{sec:CLON}
Another possible eavesdropping alternative for Eve is not to
reduce the number of photons in the signal as in the \PNS\ attack.
Instead, she can interact with the signals via a photon-number
conserving interaction of a probe system with the signal photons.
Then, after the information about the polarization basis is
publicly revealed, Eve can obtain information about the key by
measuring her probe. In this attack, multi-photon signals maintain
their photon-number and can therefore contribute with higher
probability to a "click" event. Therefore,  the fraction of the
single-photon signals in this attack can be decreased. In
principle, one would like to optimize this type of attack over all
possible probes and their interaction.  However, for the sake of
simplicity, we restrict our analysis to the case of two particular
interactions representing cloning machines. This is motivated by the fact that the optimal individual attack for
single-photon pulses coincides with the optimal phase-covariant
cloning machine \cite{Bruss00}. These
cases already proof our point.

\begin{figure}
\begin{center}
\includegraphics[scale=.7]{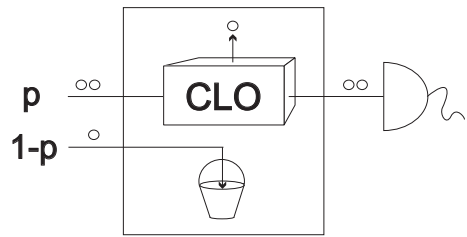}
\end{center}
\caption{When the pulse contains two photons, Eve employs an
asymmetric cloning machine which produces three clones. She keeps
one of the clones, and she sends the other two clones to Bob. This
occur with probability $p$. In the case of a single-photon pulse
(probability $1-p$), she blocks it.\label{CLO}}
\end{figure}

Consider the process represented in Fig.~\ref{CLO}.
The input signals are the same of the process of Fig.~\ref{PNS}.
But here Eve employs an asymmetric cloning machine for all
two-photon pulses, while she blocks all the single-photon pulses.
The parameter $p$ can be selected such that the processes of
Fig.~\ref{PNS} and Fig.~\ref{CLO} have the same raw key rate.
This is done in Sec.~\ref{sec:COMP}, when we compare them. We
consider two particular asymmetric cloning machines that have
been proposed by Ac\' \i n {\it et al.} in \cite{Acin03}. They
generalize the $1\rightarrow{}2$ asymmetric
cloning machines introduced in \cite{Cerf00,Niu98} to the
$2\rightarrow{}3$ case. But before introducing these cloning
machines and study the performance of the process of
Fig.~\ref{CLO} for each of them (strategies A and B below), we
introduce a qubit representation for the two-photon pulses
emitted by Alice that is used in the next subsections.

The set of two-photon signals employed in the BB$84$ protocol
span a three dimensional Hilbert space. They can be represented in
the {\it symmetric} subspace of two-qubits, which contains the
signal states $\ket{0}^{\otimes{}2}, \ket{1}^{\otimes{}2},
\ket{+}^{\otimes{}2}$, and $\ket{-}^{\otimes{}2}$, where
$\ket{\pm}=1/\sqrt{2}(\ket{0}\pm\ket{1})$.

\subsection{STRATEGY A}
In this attack Eve uses an asymmetric universal cloning machine.
It takes as an input state two copies of an unknown one-qubit
state, plus a two-qubit probe. Its unitary transformation is
defined by \cite{Acin03}
\begin{eqnarray}
    U\ket{\psi}^{\otimes 2}\ket{00}&=&\alpha\ket{\psi}^{\otimes 2}
    \ket{\phi^+}+
    \beta\Big(\tilde\si_z\ket{\psi}^{\otimes 2}\ket{\phi^-}+\nonumber\\
    &&\tilde\si_x\ket{\psi}^{\otimes 2}\ket{\psi^+}+
    i\tilde\si_y\ket{\psi}^{\otimes 2}\ket{\psi^-}\Big) ,
\end{eqnarray}
where the operator $\tilde\si_k=\si_k\otimes\one+\one\otimes\si_k$
(for $k=x,y,z$) with the usual Pauli operators $\sigma_k$, the
states $\ket{\psi^{\pm}}=1/\sqrt{2}(\ket{01}\pm\ket{10})$,
$\ket{\phi^{\pm}}=1/\sqrt{2}(\ket{00}\pm\ket{11})$, and
$\alpha^2+8\beta^2=1$. In the output, the state of the first two
qubits belong to the symmetric subspace of two-qubits signals,
and correspond to the two photons that are sent to Bob. The third
and fourth qubit constitute the probe that is kept by Eve. Next
we calculate the information that Eve can obtain on Alice's
signal as part of a sifted key by measuring her probe after the
public announcement of basis.

Eve's probe for the signals $\ket{+}^{\otimes{}2}$, and
$\ket{-}^{\otimes{}2}$, after applying the cloning machine, is
given by
\begin{equation}
\rho_{+}=2D\ket{-+}\bra{-+}+(1-2D)\ket{\varphi_{+}}\bra{\varphi_{+}},
\end{equation}
and
\begin{equation}
\rho_{-}=2D\ket{+-}\bra{+-}+(1-2D)\ket{\varphi_{-}}\bra{\varphi_{-}},
\end{equation}
respectively, where
$\ket{\varphi_{\pm}}=1/\sqrt{1-2D}(\sqrt{1-4D}\ket{\phi^{+}}\pm\sqrt{2D}\ket{\psi^{+}})$
\cite{Note4}. Note that in the subspace spanned by the two-qubit
states $\ket{-+}$ and $\ket{+-}$ Eve can discriminate between
$\rho_{+}$ and $\rho_{-}$ perfectly. In the orthogonal subspace
spanned by $\ket{\phi^{+}}$ and $\ket{\psi^{+}}$, however, the
states $\rho_{+}$ and $\rho_{-}$ present a non-vanishing overlap
$x=\langle\varphi_{+}|\varphi_{-}\rangle=(1-6D)/(1-2D)$. In this
subspace, Eve's maximum information is given by
$I_{AE}=1/2\Phi(\sqrt{1-x^2})$ \cite{Levitin95,Fuchs98}. This
means that Eve's maximum information in this cloning machine can
be written as a function of $D$ as
\begin{equation}\label{I_CLON1}
I_{AE}^{A}=2D+\frac{1-2D}{2}\Phi\bigg(\frac{\sqrt{8D(1-4D)}}{1-2D}\bigg)\;.
\end{equation}
This expressions holds also for the signals of the other
polarization basis, so that it denotes also the total Shannon
information over all signals.

Here, and also in the next subsection, we consider that double
click events are not discard by Bob, but they contribute to the
raw key. Every time Bob obtains a double click, he just decides
randomly the bit value \cite{Norbert99}.

\subsection{STRATEGY B}
The second cloning machine we consider is a phase-covariant
cloning machine. The unitary transformation of this cloning
machine is given by \cite{Acin03}
\begin{equation}
    U\ket{\varphi}\ket{00}=(V\ket{\varphi}\ket{0})
    \ket{0}+(\tilde V\ket{\varphi}\ket{0})\ket{1} ,
\end{equation}
where
$\ket{\varphi}$ can be any state in the symmetric 2-qubit Hilbert
space, and $V$ is the unitary transformation
\begin{eqnarray}\label{V}
    V\ket{00}\ket{0}&=&\ket{000} \\
    V\ket{\psi^+}\ket{0}&=&\frac{\cos\gamma(\ket{010}+
    \ket{100})+\sin\gamma\ket{001}}{\sqrt{1+\cos^2\gamma}} \nonumber\\
    V\ket{11}\ket{0}&=&\frac{\cos\gamma\ket{110}+
    \sin\gamma(\ket{011}+\ket{101})}{\sqrt{1+\sin^2\gamma}}\nonumber,
\end{eqnarray}
$\tilde V$ has the same form as $V$ but interchanging zeros and
ones on the right hand side of Eq.~(\ref{V}), and
$0\leq\gamma\leq\pi$. The first two output qubits of this cloning
machine belong again to the symmetric subspace of two-qubits
signals and correspond with the two photons which are sent to
Bob, while the other two qubits constitute Eve's probe.

Following the same argumentation used in strategy A, when Alice
sends the signals $\ket{+}^{\otimes{}2}$, $\ket{-}^{\otimes{}2}$
\cite{Note5}, the state of Eve's probe is given by $\rho_{+}$ or
$\rho_{-}$, depending on the particular state chose by Alice. The
states $\rho_{+}$ and $\rho_{-}$ can be written in the basis
$\ket{+}\ket{+}$, $\ket{+}\ket{-}$, $\ket{-}\ket{+}$,
$\ket{-}\ket{-}$ as
\begin{equation}\label{Coefa}
\rho_{+}=\frac{1}{16}\left(
\begin{array}{cccc}
a & 0 & 0 & b\\
0 & d & e & 0\\
0 & e & f & 0\\
b & 0 & 0 & c\\
\end{array}\right),\ \rm{and}\ \,\rho_{-}
=\frac{1}{16}\left(
\begin{array}{cccc}
c & 0 & 0 & b\\
0 & f & e & 0\\
0 & e & d & 0\\
b & 0 & 0 & a\\
\end{array}\right),
\end{equation}
respectively, where the coefficients $a, b, c, d, e$, and $f$ are
complicated functions of the parameter $\gamma$. The exact
expression of these coefficients is given in appendix B. To
calculate Eve's maximum information in this case, we can decompose
Eve's optimal measurement again in two steps. First, she performs
a projection measurement onto the two orthogonal subspaces spanned
by $\ket{+}\ket{+}$, $\ket{-}\ket{-}$, and $\ket{+}\ket{-}$,
$\ket{-}\ket{+}$, respectively. This measurement reduces the
optimization problem in the whole space, to discriminate in each
subspace between two equiprobable one-qubit states whose density
matrices have the same invariants. This problem was solved by
Levitin in \cite{Levitin95}. The maximum of the mutual information
in each subspace is given by $I=1/2\Phi(\sqrt{1-r-2d})$, where $r$
represents the trace of the product of the two states, and $d$ is
the determinant of their density matrices. Using the expressions
for Eve's maximum information in each subspaces, one can obtain
Eve's maximum information in the cloning machine as a function of
the coefficients $a, c, d$, and $f$. It is given implicitly by
\begin{equation}\label{I_CLON2}
I_{AE}^{B}=\frac{1}{32}\bigg\{(a+c)\Phi\bigg(\frac{a-c}{a+c}\bigg)+(d+f)\Phi\bigg(\frac{d-f}{d+f}\bigg)\bigg\}.
\end{equation}
The disturbance $D$ in this case has the form
\begin{equation}
D=\frac{1}{2}\bigg\{1-\frac{1}{\sqrt{2(1+\cos^2\gamma)}}\big(\cos\gamma+\frac{1}{\sqrt{1+\sin^2\gamma}}\big)\bigg\}.
\end{equation}
Again, due to symmetry with respect to the polarization bases,
Eq.~(\ref{I_CLON2}) holds also for the total average Shannon
information.

\section{PNS ATTACK VERSUS CLONING ATTACKS}\label{sec:COMP}
\begin{figure}
\begin{center}
\includegraphics[scale=.3]{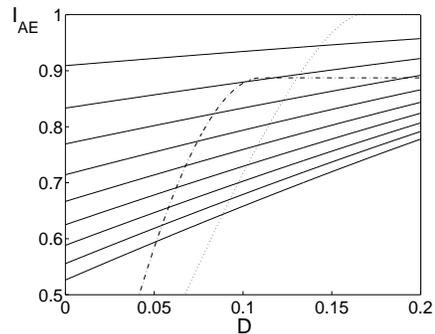}
\end{center}
\caption{Eve's maximum information versus the disturbance D: \PNS\
process for increasing, equally spaced values of $\eta_{det}$
(solid). The lower line corresponds to $\eta_{det}=0.1$, while the
upper line corresponds to $\eta_{det}=0.9$. Universal cloning
machine, Strategy A (dotted). Phase-covariant cloning machine,
Strategy B (dashdot).\label{RES}}
\end{figure}

The processes represented in Fig.~\ref{PNS} and in Fig.~\ref{CLO},
for both cloning machines, gives a symmetric detection pattern.
That is, if Bob measures the signals in the same basis chosen by
Alice when preparing the states, then the probability of obtaining
a correct result, a wrong result, or a double click is the same
for all the signals. Otherwise the outcomes corresponding to
events with one detection click are completely random. For a fair
comparison of the \PNS\ process and the two cloning processes, we
need to assure that the raw bit rate in Bob's detectors is the
same for all of them,
$p\eta_{det}+(1-p)\eta_{det}=p\eta_{det}(2-\eta_{det})$. The
left-hand-side of the equation is the number of clicks of the
\PNS\ process, while the right-hand-side is the number of clicks
expected in Fig.~\ref{CLO} for both cloning machines. This means
that $p=1/(2-\eta_{det})$. If we include this value in
Eq.~(\ref{eqPNS}), Eve's maximum information in the \PNS\ process
is now written as
\begin{equation}
I_{AE}^{PNS}=\frac{1}{2-\eta_{det}}\bigg\{1+\frac{1-\eta_{det}}{2}\Phi\Big(2\sqrt{D(1-D)}\Big)\bigg\}.
\end{equation}
This expression can now be directly compared with
Eq.~(\ref{I_CLON1}) and Eq.~(\ref{I_CLON2}). The results are
plotted in Fig.~\ref{RES} and show regimes of $D$ for which
the process based on cloning machines provides Eve with more
information than the \PNS\ process. Note that Eve's maximum
information in the cloning processes is independent of
$\eta_{det}$. This fact comes from the matching condition for the
raw bit rate. In the \PNS\ process, as expected, when
$\eta_{det}$ approaches one, also $I_{AE}^{PNS}$ approaches one,
since the \PNS\ attack is the optimal strategy for Eve's in the
case of ideal detectors. In the phase-covariant cloning machine
of strategy B, Eve's maximum information never reaches one. The
reason is that in this particular cloning machine none of the two
qubits kept by Eve can reach a fidelity one with respect to the
input state. For low values of $D$, this cloning machine gives
Eve more information than the universal cloning machine of
strategy A. From the perspective of cloning machines this fact is
not surprising. The fidelity achievable in the clones depends
always on the set of allowed input states. As more information
about the input set is known, better the input states can be
cloned. The phase covariant cloning machine exploits the fact
that the input states are {\it equatorial} qubits. That is, the
$z$ component of their Bloch vector is zero. The cloning machine
of strategy A, however, is designed to clone {\it any } input
qubit with the same fidelity.
\subsection{OBSERVED ERROR RATE}
\label{observed} In this section we obtain the relationship
between the disturbance $D$, which appears in Fig.~\ref{RES}, and
the overall observed error rate $e$ which is measured in the
experiment. This relationship can be established by an analysis
of the \PNS\ attack alone, which includes here the optimal
eavesdropping on single photon signals, as before.

\begin{figure}
\begin{center}
\includegraphics[scale=.3]{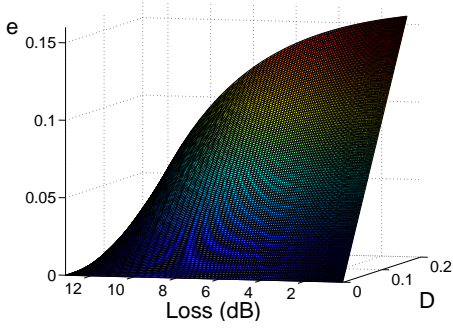}
\end{center}
\caption{Observed error rate $e$ versus the disturbance $D$ as a
function of the loss in dB of the quantum channel. The mean photon
number $\mu$ is $0.1$ and $\eta_{det}$ is $0.2$ in this
example.\label{figeD}}
\end{figure}

The probability that a multi-photon signal undergoes the \PNS\
attack and then is detected by Bob's detection device is given by
\begin{equation}\label{arr}
P_{arr}^{multi}=\sum_{n=2}^{\infty} P(n,\mu)[1-\bar{\eta}^{n-1}],
\end{equation}
where $P(n,\mu)=e^{-\mu}\mu^n/n!$ is the photon-number
distribution of the signal states emitted by Alice, given by
Eq.~(\ref{SigAlice}). On the other hand, the expected click rate
at Bob's side has the form
\begin{equation}\label{exp}
P_{exp}=1-e^{-\mu\eta_{det}\eta_{t}},
\end{equation}
where $\eta_{t}$ is the transmission efficiency of the quantum
channel. The total loss in $dB$ of the quantum channel is given by
$-10\log_{10}\eta_{t}$. From $P_{arr}^{single}$ and $P_{exp}$ one
can obtain the probability that single-photon signals contribute
to the raw key. It is given by
\begin{equation}\label{sing}
P_{arr}^{single}=P_{exp}-P_{arr}^{multi}.
\end{equation}
The multi-photon pulses does not introduce any error in the
sifted key. This means that
\begin{equation}\label{eD}
e=\frac{P_{arr}^{single}}{P_{exp}}D
\end{equation}
After substituting the values of $P_{exp}$ and $P_{arr}^{single}$
into Eq.~(\ref{eD}) we finally obtain
\begin{equation}\label{eD2}
e=\frac{e^{-\mu}(\eta_{det}e^{\mu\eta_{det}\eta_{t}}+e^{\mu}(1-\eta_{det})-e^{\mu(1-\eta_{det}(1-\eta_{t}))})}{(1-\eta_{det})(1-e^{\mu\eta_{det}\eta_{t}})}D
\end{equation}
This result is illustrated in Fig.~\ref{figeD} for some typical
values of $\mu$ and $\eta_{det}$.
 When the loss in the channel increases, the observed error rate
$e$ for each value of $D$, as expected, decreases. The regime that
we consider here corresponds with a value of the loss in the
channel such as Eve can perform a \PNS\ attack on all the
multi-photon pulses, but $P_{exp}>P_{arr}^{multi}$. The first
condition requires
\begin{equation}
P_{exp}\leq{}\eta_{det}P(1,\mu)+P_{arr}^{multi},
\end{equation}
which provides an upper bound for the transmission efficiency of
the channel
\begin{equation}\label{int1}
\eta_{t}\leq{}-\frac{\ln[\frac{e^{-\mu}(e^{\mu(1-\eta_{det})}-\eta_{det}(1+\mu(1-\eta_{det})))}{1-\eta_{det}}]}{\mu\eta_{det}}
\end{equation}
The second constraint $P_{exp}>P_{arr}^{multi}$ implies a lower
bound for $\eta_{t}$
\begin{equation}\label{int2}
\eta_{t}>-\frac{\ln[\frac{e^{-\mu\eta_{det}}-\eta_{det}e^{-\mu}}{1-\eta{det}}]}{\mu\eta_{det}}
\end{equation}
The meaning of this condition is to guarantee that we are not in a
regimen where the \PNS\ attack is still Eve's optimal strategy.
When $\mu=0.1$ and $\eta_{det}=0.2$, we obtain for the lower and
upper bound  $0.17{\rm dB}$ and $13.2 {\rm dB}$, respectively.

\begin{figure}
\begin{center}
\includegraphics[scale=.3]{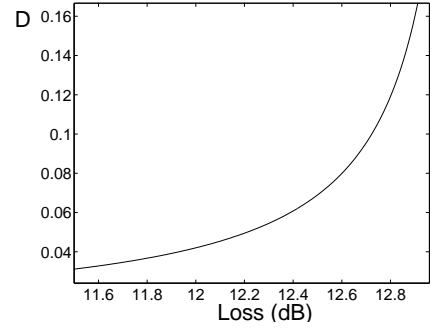}
\end{center}
\caption{The disturbance $D$ as a function of the loss in dB of
the quantum channel for a fix value of the observed error rate
$e=0.01$. The mean photon number $\mu$ is $0.1$ and $\eta_{det}$
is $0.2$. \label{figuno}}
\end{figure}

The process based on cloning machines become more powerful than
the \PNS\ process for lower values $e$ when the loss is high. The
typical value of the observed error rate in the experiments is
around $1\%$ if we consider only errors in the quantum channel.
Therefore, it is interesting to see how the value of the
disturbance $D$ change as a function of the loss in the channel
when we impose $e$ to be $1\%$. This is illustrated in
Fig.~\ref{figuno}. We find, in combination with Fig. \ref{RES}
that for losses higher than $12.5 {\rm dB}$ the \PNS\ attack is
clearly no longer optimal for these typical parameters.

It is worth to point out that when the losses in the channel are
small, but still inside the interval imposed by Eq.~(\ref{int1})
and Eq.~(\ref{int2}), the eavesdropping attack which includes the
cloning machine can be made more powerful than the \PNS\ attack
even for a lower value of $e$ than the one given in
Eq.~(\ref{eD2}). The reason is that, although in this situation
Eve cannot discard too many single-photon pulses, she can
re-distribute the errors from the single-photon processes into
the two-photon processes. To this end, she increases her
intrusion via the cloning attack on the two-photon signals, while
she reduces her intrusion on the single-photon signals. The
exploitation of this effect is beyond the scope of this article.
\section{PHOTON STATISTICS}\label{sec:STAT}
In the previous sections we consider the case of the standard
BB$84$ protocol, where only the raw bit rate is monitored. Here,
we briefly discuss the case of the extended version of the
protocol, where Alice and Bob uses the full statistics at their
disposal to detect Eve.

In this scenario, it is straightforward to see that the processes
of Fig.~\ref{PNS} and Fig.~\ref{CLO} are not equivalent. The \PNS\
process of Fig.~\ref{PNS} can never produce a double click in
Bob's detectors, while the process of Fig.~\ref{CLO} presents
always a non-vanishing probability of producing a double click,
independently of the basis that Bob uses for his measurement. In
fact, the \PNS\ attack never produces a double click event when
Bob chooses for his measurement the same basis that Alice used
when preparing the signals. This means that, in principle, Alice
and Bob might employ this information to discard any eavesdropping
strategy that includes the cloning process. However, if we
consider a real implementation of the protocol, then the situation
is not so simple. The reason is that the quantum channel is not
just lossy, but presents a misalignment that introduces errors in
the signals \cite{Gisin01}. As a result we have that any
multi-photon signal has a non-zero probability of providing a
double click, independently of the basis used by Bob in his
measurement. This means that Eve must adapt the \PNS\ attack such
that it reproduces the expected misalignment in the channel.
Otherwise her attack would be detected. In particular, Eve have to
introduce some noise in the signals that are sent to Bob. In the
case of single-photon pulses, this can be achieved by sending the
signals through a depolarizing channel of appropriate parameters.
This is precisely the effect of the symmetric \OA\ introduced in
Sec.~\ref{sec:PNS}. Therefore, in these pulses, Eve can always get
information from this extra noise. The multi-photon pulses,
however, gives already Eve full information about the key, and she
cannot exploit the noise she needs to introduce to get more
information from the single-photon pulses.

The eavesdropping attack which includes the cloning machine can
also be adapted such that it reproduces the statistics that is
expected from a realistic channel. However, the question whether
 it remains more powerful than the \PNS\ attack, in this scenario,
requires a deeper analysis. If we consider the situation where Eve
performs a \PNS\ attack on the pulses that contain more than two
photons, and the misalignment in the channel is sufficiently
strong for Eve to get full information from the cloning process,
then Eve can obtain as much information as with the \PNS\ attack.
If the misalignment is smaller, it seems that the strategy that
combines the cloning process with the \PNS\ attack on the remaining
multi-photon pulses cannot be more powerful than the \PNS\ attack.
The reason is that the adapted version of the \PNS\ attack still
contains processes that do not produce any double click in Bob's
detectors, independently of the basis that Bob uses for his
measurement. To compensate this effect, Eve has to subtract more
than one photon from the multi-photon pulses, such that she
creates processes that do not produce double clicks. But now the
effectiveness of the complete strategy decreases, since the
probability that the signals which provide Eve full information
about the key (multi-photon pulses) contribute to the raw key
decreases.

Although this fact constitutes a handicap of the eavesdropping
strategy that combines the cloning process of Fig.~\ref{CLO} with
the \PNS\ attack on the rest of the pulses, it might be of
relative importance in practice. Double clicks are rare events
that have a very small probability to occur, and the statistical
fluctuations in the channel, together with the effect of dark
counts in Bob's detectors, makes the detection of Eve's presence
not easy. Moreover, Eve might also use a mixed strategy that
combines probabilistically the \PNS\ process of Fig.~\ref{PNS} and
the cloning process of Fig.~\ref{CLO}, such as her attack remains
still more powerful than the \PNS\ attack, while making her
detection even more difficult.
\section{CONCLUSION}\label{sec:CONC}
In an ideal quantum optical implementation of \QKD, the sender use
single photons to encode the information he transmits. However,
current experiments are not based on single photon sources, but
they are usually based on weak coherent pulses (WCP) with a low
average photon number. Also the detectors employed by the receiver
are not perfect, but have a low detection efficiency and are
noisy. This fact, together with the loss in the quantum channel,
limits the distances that can be covered by these methods. In this
scenario, it is tempting to assume that the loss in the detectors
cannot be changed by Eve in order to increase the covered
distance, while the \PNS\ attack, like in the case of a
conservative definition of security, still constitutes Eve's
optimal strategy. In this paper we disprove this belief for the
case of the standard BB$84$ protocol, where only the raw bit rate
(before the key distillation phase) is monitored. We constructed
two specific eavesdropping strategies which include processes that
do not subtract photons from the pulses, and that are more
powerful than the \PNS\ attack for some relevant regimes of the
observed error rate. These strategies are based on the use of
cloning machines. A complete analysis of Eve's optimal attack in
this situation is still missing. In the extended version of the
BB$84$ protocol, where Alice and Bob consider the full statistics
at their disposal, the situation is not as straightforward, and a
deeper security analysis of this scenario is required.
\section{ACKNOWLEDGEMENTS}
The authors wish to thank A. Ac\'\i n, A. Dolinska, P. van Loock
and Ph. Raynal for very useful discussions, and especially K.
Tamaki for his critical discussion of this article. We also like
to thank Gerald Gilbert for the clarifying discussion of his
investigations. This work is supported by the DFG under the Emmy
Noether programme and the network of competence QIP of the state
of Bavaria.

\appendix
\section{Related work}
\label{why} An investigation of the scenario where Eve cannot
improve Bob's detectors has been undertaken in \cite{gilbert00a}.
We believe that this investigation is incomplete so far. The
authors claim that they have performed an thorough analysis of
the situation where Bob's detection efficiency cannot manipulated
by Alice, together with the restriction of an 'individual
attack'. The authors define 'individual attack' radically
different from the usual terminology that is used in the analysis
of quantum key distribution: they refer to attacks that act on
{\em photons} individually, rather than {\em signals}. In practice
this means that the authors consider optimal attacks on single
photon signals, which can be implemented by attaching a probe to
a single photon, on the other hand, they disallow attaching a
probe to a two-photon signal, since that would mean to interact
with the two photons 'coherently'. In fact, they state
\cite{gilbert00a} that such manipulations would be possible only
when quantum computers become available.

Of course, it is not unusual to start with some assumptions about
restrictions of eavesdropping strategies. For example,
investigation of an individual attack scenario (now referring to
the standard definition that relates to the signal pulses) has
proven to be  very powerful since the analysis can be performed
easily  and the resulting parameters for privacy
amplification and the  secure key rate corresponds
roughly to the subsequently derived values that assure security
again all attacks, including coherent attacks on all signals. From
this experience the individual attack  derives its role as a first
step investigation of the performance and security analysis of
\QKD\ schemes. The relationship between individual and coherent
attacks has been strengthened by the results of Wang
\cite{Xiangbin01}.

In the scenario considered by  \cite{gilbert00a} we cannot see an
equivalent role. Another motivation to investigate restricted
scenarios might be the technological challenge of different
eavesdropping strategies. However,  the technological
difference between attaching a probe to a single photon as
compared to attaching a probe to a two-photon signal is not
evident. Clearly, these questions do not invalidate the obtained
results. However, in our point of view, the authors of
\cite{gilbert00a} are inconsistent in describing the restrictions
of their considered eavesdropping attacks. They claim that as a
consequence of their restriction they need to consider only three
types of attacks:
\begin{enumerate}
\item
'Direct attacks' in which Eve can unambiguously determine the
signal state by a direct measurement of the signal. This
corresponds to the unambiguous state discrimination attack in
\cite{dusek00a}. (Requires at least $3$ photons.)
\item 'Indirect attacks', which is precisely the \PNS\ attack
\cite{Huttner95,Norbert00,Norbert00b} that extract one photon from
the signal. (Requires at least $2$ photons.)
\item 'Combined attacks' which performs the indirect and the
direct attack. (Requires therefore at least $5$ photons, and is in
the analysis later on shown to be an inferior attack.)
\end{enumerate}
It is left open how these categories emerge and why this should be
a complete description. As first point of criticism note that the
authors apply for the  'direct attack' the results of
\cite{dusek00a} that provide the performance for optimal
unambiguous state discrimination measurements. Can we implement
this attack by acting 'individually' on photons?  A second point
of criticism is that to perform these attacks Eve needs to know
the number of photons in each pulse. If one thinks of photons as
distinguishable particles in a pulse, this might be easy. In a
proper quantum optical description, however, these type of
counting mechanisms, which do not disturb the signal, will require
in all experience the same level of interaction between a probe
and the total signal, as does a general eavesdropping attack on
the signal.

So far, we  pointed at inconsistencies that do not endanger the
security statement derived in \cite{gilbert00a}. These attacks
overestimate Eve's capabilities as compared to the initial
restriction that require 'individual' attacks on photons. However,
the categorization by Gilbert and Hamrick  left out possible
attacks. Those attacks still operate on 'individual' photons only.
As an example let us consider a two-photon pulse. According to
\cite{gilbert00a} the only attack we need to consider is the \PNS\
attack. Instead, let Eve perform a direct measurement on the
photons, for example in the sense of a minimum-error measurement.
Of course, the error will be non-zero, since on a two-photon state
in the BB84 polarizations one cannot perform successfully
unambiguous state discrimination. However, optimal eavesdropping
on single-photon signals also results in some errors.   Another
attack would be to separate the two photons. Then one can attack
probes to both photons and try to combine the photons again in
Bob's detection apparatus, e.g. by sending them to Bob in close
sequence so that Bob does not notice that they have been
separated. Moreover, similar attacks are omitted for higher
photon numbers.

These examples question the completeness of the proposed
classification of eavesdropping attacks in \cite{gilbert00a}.
Note that after receiving an advance copy of this manuscript the
authors of \cite{gilbert00a} revised their work, acknowledging
the incompleteness of their analysis. This means that we have to
treat the classification as an assumption that only those three
classes are of relevance. This includes the assumption that for
the two-photon pulse the \PNS\ attack is optimal in their
restricted scenario. As a consequence, a security claim for an
experimental implementation of \QKD\ should not be based on this
analysis, as done in \cite{gilbert00a,Gilbert2}, since it
underestimates Eve's ability. Nevertheless, within the three
investigated classes of eavesdropping attacks, Gilbert and
Hamrick have been able to show that the unambiguous state
discrimination attack can be more effective for Eve than the
photon-number splitting attack for signals containing three or
more photons.

\section{Explicit Expressions}
In this appendix we provide the exact expressions for the
coefficients $a, b, c, d, e$, and $f$ that are introduced in
Eq.~(\ref{Coefa}), as a function of the angle $\gamma$.
\begin{eqnarray*}
a&=&1+\frac{4\sin\gamma}{\sqrt{3+\cos(2\gamma)}}+\frac{10\sin(2\gamma)}{\sqrt{3+\cos(2\gamma)}\sqrt{1+\sin^2\gamma}}\nonumber\\
&&+\frac{2\cos\gamma}{\sqrt{1+\sin^2\gamma}}+\cos^2\gamma\bigg(\frac{8}{1+\cos^2\gamma}+\frac{1}{1+\sin^2\gamma}\bigg)\nonumber\\
&&+4\sin^2\gamma\bigg(\frac{1}{3+\cos(2\gamma)}+\frac{1}{1+\sin^2\gamma}\bigg),\nonumber
\end{eqnarray*}
\begin{eqnarray*}
b&=&1+\frac{2\cos\gamma}{\sqrt{1+\sin^2\gamma}}+\frac{8\sin^2\gamma(9+\cos(2\gamma))}{-17+\cos(4\gamma)}\qquad\qquad\nonumber\\
&&+\cos^2\gamma\bigg(\frac{8}{1+\cos^2\gamma}+\frac{1}{1+\sin^2\gamma}\bigg),\nonumber
\end{eqnarray*}
\begin{eqnarray*}
c&=&1-\frac{4\sin\gamma}{\sqrt{3+\cos(2\gamma)}}-\frac{10\sin(2\gamma)}{\sqrt{3+\cos(2\gamma)}\sqrt{1+\sin^2\gamma}}\nonumber\\
&&+\frac{2\cos\gamma}{\sqrt{1+\sin^2\gamma}}+\cos^2\gamma\bigg(\frac{8}{1+\cos^2\gamma}+\frac{1}{1+\sin^2\gamma}\bigg)\nonumber\\
&&+4\sin^2\gamma\bigg(\frac{1}{3+\cos(2\gamma)}+\frac{1}{1+\sin^2\gamma}\bigg),\nonumber\\
\end{eqnarray*}
\begin{eqnarray*}
d&=&1+\frac{4\sin^2\gamma}{3+\cos(2\gamma)}+\frac{\cos^2\gamma}{1+\sin^2\gamma}-\frac{2\cos\gamma}{\sqrt{1+\sin^2\gamma}}\nonumber\\
&&+\frac{4\sin\gamma(-\cos\gamma+\sqrt{1+\sin^2\gamma})}{\sqrt{3+\cos(2\gamma)}\sqrt{1+\sin^2\gamma}},\nonumber
\end{eqnarray*}
\begin{eqnarray*}
e&=&1-\frac{4\sin^2\gamma}{3+\cos(2\gamma)}+\frac{\cos^2\gamma}{1+\sin^2\gamma}-\frac{2\cos\gamma}{\sqrt{1+\sin^2\gamma}},\quad\nonumber\\
\rm{and}
\end{eqnarray*}
\begin{eqnarray*}
f&=&1-\frac{4\sin\gamma}{\sqrt{3+\cos(2\gamma)}}+\frac{4\sin^2\gamma}{3+\cos(2\gamma)}+\frac{\cos^2\gamma}{1+\sin^2\gamma}\nonumber\\
&&-\frac{2\cos\gamma}{\sqrt{1+\sin^2\gamma}}+\frac{2\sin(2\gamma)}{\sqrt{3+\cos(2\gamma)}\sqrt{1+\sin^2\gamma}},\nonumber
\end{eqnarray*}

\bibliographystyle{apsrev}

\end{document}